\documentstyle[preprint,aps]{revtex}
\tightenlines
\begin{document}
\title{Scheme to Measure Quantum Stokes Parameters and their Fluctuations and
Correlations}
\author{G. S. Agarwal\thanks{email: gsa@prl.ernet.in}}
\address{Physical Research Laboratory, Navrangpura, Ahmedabad 380009, India}
\author{S. Chaturvedi\thanks{email: scsp@uohyd.ernet.in}}
\address{ School of Physics, University of Hyderabad, Hyderabad 500046,
India}
\date{\today}
\maketitle
%\draft{16.8.2000)
\begin{abstract}
We propose a scheme to measure quantum Stokes parameters, their fluctuations 
and correlations. The proposal involves measurements of intensities and 
intensity-intensity correlations for suitably defined 
modes, which can be 
produced by a combination of half wave and quarter wave plates.
\end{abstract}
\maketitle
\newpage
Polarisation of light is a well studied subject. Text book \cite{born}
descriptions of light polarisation are based on the Stokes parameters. 
Consider, for example, a light beam propagating in $z$ direction and 
decomposed into two components polarised along the orthogonal directions 
$\hat{\epsilon}_i, i=1,2$:
\begin{equation}
\vec{E}(z,t)\equiv(\hat{\epsilon}_1 \alpha_1 + \hat{\epsilon}_2 \alpha_2 )
 e^{ikz-iwt} + ~~ c.c. . 
\end{equation}
The four Stokes parameters characterising such a beam are  defined 
by \cite{born}
\begin{eqnarray}
S_0 &=& |\alpha_1|^2 + |\alpha_2 |^2 ~~~~;~~~~  S_1 = |\alpha_1 |^2 - |\alpha_2 |^2~~;
\nonumber \\
S_2 &=& (\alpha_1^* \alpha_2 + \alpha_1 \alpha_2^*)~~~;~~~  S_3  =  -i 
(\alpha_1^* \alpha_2 - \alpha_1 \alpha_2^*),
\end{eqnarray}
which, as is well known, lead to the representation of light polarisation by 
a point on the Poincar\'e sphere. All states of linear polarisation lie in 
the $xy$ plane, whereas north and south poles represent states of circular 
polarisation. All other points on Poincar\'e sphere represent states of 
elliptic polarisation. Methods to measure all four Stokes parameters are also 
well known. 

The Stokes parameters have been
generalised in several ways\cite{mandel,wolf,korol,chirkin,abouraddy,Jauch}.
For example Wolf \cite{wolf} has shown how the polarisation characteristics 
of partially coherent light can be calculated. In such a case one has to 
average quantities defined by (2), over an ensemble distribution of 
$\alpha_1$ and$\alpha_2$. More recently, there has been considerable 
interest in generalising these ideas to discuss the polarisation
characteristics of a non classical field possessing very strong quantum
fluctuations \cite{korol,chirkin}. In such a case the field 
$\vec{E}$ in (1) becomes an operator with $\alpha_1$ and $\alpha_2$ replaced 
by the annihilation operators $a_1$ and $a_2$. We can then define quantum 
Stokes operators  \cite{Jauch}, in analogy to (2), as :
\vskip0.5cm
\begin{mathletters}
\begin{eqnarray}
{\cal S}_0 &=& (a_{1}^{\dagger}a_{1} + a_{2}^{\dagger}a_2),\\
{\cal S}_1 &=& (a_{1}^{\dagger}a_{1} - a_{2}^{\dagger}a_2),\\
{\cal S}_2 &=& (a_{1}^{\dagger}a_{2} + a_{2}^{\dagger}a_1),\\
{\cal S}_3 &=& -i(a_{1}^{\dagger}a_{2} - a_{2}^{\dagger}a_1)~;~
[a_i,a_j^{\dagger}]=\delta_{ij}.
\end{eqnarray}
\end{mathletters}
The expectation values $\langle{\cal S}_i\rangle~,~i=0,1,2,3$ can be identified with the 
Stokes parameters $S_{i}~,~i=0,1,2,3$. \\
As is well known, the above  operators furnish the Schwinger representation 
\cite{schwinger} of the $SU(2)$ algebra
\begin{mathletters}
\begin{eqnarray}
\left[{\cal S}_i, {\cal S}_j\right] &=& 2i\epsilon_{ijk}{\cal S}_k,~i,j,k=1,2,3 \\ 
 \left[{\cal S}_0, {\cal S}_i\right] &=& 0,~i=1,2,3 \\
{\cal S}_{0}({\cal S}_{0}+2)&=&{\cal S}_{1}^{2}+{\cal S}_{2}^{2}+
{\cal S}_{3}^{2}.
\end{eqnarray}
\end{mathletters}
The connection to the $SU(2)$ algebra is very useful and  many of the concepts
developed in connection with spin squeezing \cite{agarwal} have been taken
over to discuss polarisation squeezing \cite{korol,chirkin,grangier,puri}.
The Stokes parameters $S_i~, i= 0,1,2,3$ involve four averages 
$\langle a_{1}^{\dagger}a_{1}\rangle,~\langle a_{2}^{\dagger}a_{2}\rangle,~
\langle a_{1}^{\dagger}a_{2}\rangle,~ 
\langle a_{2}^{\dagger}a_{1}\rangle$. Further, the variances 
\begin{equation}
V_{ij}=\frac{1}{2}(\langle\{{\cal S}_i, 
{\cal S}_j\}\rangle-\{\langle{\cal S}_i\rangle,\langle{\cal S}_j\rangle\})~;
~i, j = 0,\cdots,3~,i\leq j, 
\end{equation} 
can be expressed in terms of the nine normally ordered field correlations 
 $\langle a_{1}^{\dagger}a_{1}^{\dagger}a_{1}a_{1}\rangle,~ 
\langle a_{2}^{\dagger}a_{2}^{\dagger}a_{2}a_{2}\rangle,~ 
\langle a_{1}^{\dagger}a_{1}^{\dagger}a_{2}a_{2}\rangle,~ 
\langle a_{2}^{\dagger}a_{2}^{\dagger}a_{1}a_{1}\rangle,~ 
\langle a_{1}^{\dagger}a_{2}^{\dagger}a_{1}a_{2}\rangle,~ 
\langle a_{1}^{\dagger}a_{1}^{\dagger}a_{1}a_{2}\rangle,~
\langle a_{2}^{\dagger}a_{2}^{\dagger}a_{1}a_{2}\rangle,~
\langle a_{1}^{\dagger}a_{2}^{\dagger}a_{1}a_{1}\rangle,\\ 
\langle a_{1}^{\dagger}a_{2}^{\dagger}a_{2}a_{2}\rangle$.
Given the circumstance that, in an experiment, one typically measures average
intensities and intensity variances, the question one is confronted with is whether it is possible to 
deduce the relevant field correlations by measuring average intensities and 
intensity variances of suitably defined modes. The answer to this question 
is provided by a theorem due to Mukunda and Jordan \cite{mukunda} 
in a more general context. This would be
used in the following to show how all $V_{ij}$ can be essentially obtained from intensities and
intensity fluctuations. We note that very recently Korolkova {\it{et al}}
\cite{korol} have proposed a method to measure the diagonal quantities $V_{ii}$. 
An implementation of this proposal has been achieved by Bowen {\it{et al}} 
\cite {bowen}.

Before we proceed further, we note that for an important special class of the 
states of the radiation field, such as the squeezed coherent state $|\alpha_1,
\alpha_2,\zeta\rangle$, the correlations can be derived in closed analytical 
form. In this case it is well known that the Wigner function as well as the
$Q-function$ are Gaussian \cite{walls}. Thus the fourth order expectation values
can be expressed in terms of second order expectation values. If we define 
displaced modes $d_i$ = $a_i-\alpha_i $, then we have for example\\
\begin{eqnarray}
\langle d_1 d_2 d_1^{\dagger} d_2^{\dagger}\rangle &\equiv&\langle d_1
d_2\rangle\langle d_1^{\dagger} d_2^{\dagger}\rangle + \langle d_1 d_1^{\dagger}
\rangle \langle d_2 d_2^{\dagger}\rangle +
\langle d_1 d^{\dagger}_2\rangle \langle d_2 d_1^{\dagger}\rangle\\
&=& \langle d_1 d_2\rangle\langle d_1^{\dagger} d_2^{\dagger}\rangle + 
\langle d_1 d_1^{\dagger}\rangle 
\langle d_2 d_2^{\dagger}\rangle,
\end{eqnarray}
since for the squeezed state $\langle d_1 d_2^{\dagger}\rangle= 0$. It should be borne in mind
that the special case of coherent states is obtained by setting $\zeta=0$. 
Clearly if one knows, a priori,  that the field is in the squeezed state then 
all the fluctuations and correlations among Stokes parameters can be related 
to second order expectation values. For arbitrary non classical states 
characterised by non-Gaussian Wigner function, the situation would be much 
more complex. Another simple case is obtained if the fluctuations of the 
field around the mean value are small. In such a case 
$S_k - \langle S_k \rangle$ can be approximated by terms linear in 
$d_i$ and $d_i^{\dagger}$. Thus correlations $V_{ij}$ are expressible 
in terms of the expectation values of quadratic forms. The latter, in many 
cases reduces to variances in quadratures \cite{korol,bowen}.

As noted earlier, The measurement scheme proposed here is based on a  work of 
Mukunda and Jordan \cite{mukunda} which, In the present context, involves 
use of  unitary transformations defined by
\begin{equation}
\left(\matrix{
b_1 \cr b_2}\right) = 
u \left(\matrix{
a_1 \cr a_2} \right)~;~
u= \left(\matrix{ \cos\theta & e^{i\phi}\sin\theta \cr
-e^{i\phi}\sin\theta &\cos\theta }\,\right), 
\end{equation}
and measurements of the intensities 
$\langle b_i^{\dagger} b_i\rangle (i=1,2)$ and intensity correlations like 
$\langle b_1^{\dagger}b_1^{\dagger}b_1 b_2\rangle, \langle b_1^{\dagger}
b_2^{\dagger}b_1 b_2\rangle$ for different settings of $\theta,\phi$. Thus, 
for the mean intensity, we have
\begin{eqnarray}
\langle b_{1}^{\dagger}(u)b_1(u)\rangle &=& \cos^2\theta \langle a_{1}^\dagger
 a_1\rangle+\sin^2\theta \langle a_{2}^\dagger a_2\rangle\nonumber\\
&+&e^{i\phi}\sin\theta \cos\theta 
\langle a_{1}^\dagger a_2\rangle+e^{-i\phi}\sin\theta \cos\theta \langle a_{2}
^\dagger a_1\rangle.
\end{eqnarray} 
We can, therefore, determine 
$\langle a_{1}^{\dagger}a_{1}\rangle,~\langle a_{2}^{\dagger}a_{2}\rangle,~
\langle a_{1}^{\dagger}a_{2}\rangle,~ 
\langle a_{2}^{\dagger}a_{1}\rangle$ from the knowledge of 
$\langle b_{1}^{\dagger}(u)
b_{1}(u)\rangle$ , for instance, for the  four settings corresponding to 
$(\theta,\phi)$ equal to $(0,0),~(\pi/2, 0),~(\pi/4, 0),~(\pi/4,\pi/2)$. 
This is, in fact, identical to the procedure adopted for measurement of 
classical stokes parameters \cite{born}.\\

Next we turn to the nine field correlations which arise in the expressions for
the variances of the Stokes operators. On selecting the unitary operator in (8)
as 
\begin{equation}
u= \left(\matrix{
 \cos\theta & \sin\theta \cr
-\sin\theta &\cos\theta }\,\right),
\end{equation}
we find the relation
\begin{eqnarray}
\langle b_{1}^{\dagger}(u)b_{1}^\dagger(u)b_{1}(u)b_1(u)\rangle&=&
\langle a_{1}^{\dagger}a_{1}^{\dagger}a_{1}a_{1}\rangle\cos^4\theta + 
\langle a_{2}^{\dagger}a_{2}^{\dagger}a_{2}a_{2}\rangle\sin^4\theta\nonumber\\
&+&(4\langle a_{1}^{\dagger}a_{2}^{\dagger}a_{1}a_{2}\rangle+ 
\langle a_{1}^{\dagger}a_{1}^{\dagger}a_{2}a_{2}\rangle+ 
\langle a_{2}^{\dagger}a_{2}^{\dagger}a_{1}a_{1}\rangle)\cos^2\theta \sin^2\theta \nonumber
\\&+& 
2(\langle a_{1}^{\dagger}a_{1}^{\dagger}a_{1}a_{2}\rangle+
\langle a_{1}^{\dagger}a_{2}^{\dagger}a_{1}a_{1}\rangle)\cos^3\theta \sin\theta
\nonumber\\
&+&2(\langle a_{1}^{\dagger}a_{2}^{\dagger}a_{2}a_{2}\rangle+ 
\langle a_{2}^{\dagger}a_{2}^{\dagger}a_{1}a_{2}\rangle)\cos\theta\sin^3\theta,
\end{eqnarray}
which for five different values of $\theta$ allows us to determine the 
terms with different dependence on $\theta$ in terms of the measured values 
$\langle b_{1}^{\dagger}(u)b_{1}^\dagger(u)b_{1}(u)b_1(u)\rangle$ for the chosen $\theta$ 
settings. 
Next taking the unitary operator as 
\begin{equation}
u= \left(\matrix{ \cos\theta & i\sin\theta \cr
+i\sin\theta &\cos\theta }\right).
\end{equation}
we get the relation
\begin{eqnarray}
\langle b_{1}^{\dagger}(u)b_{1}^\dagger(u)b_{1}(u)b_1(u)\rangle&=&
\langle a_{1}^{\dagger}a_{1}^{\dagger}a_{1}a_{1}\rangle\cos^4\theta + 
\langle a_{2}^{\dagger}a_{2}^{\dagger}a_{2}a_{2}\rangle\sin^4\theta\nonumber\\
&+&(4\langle a_{1}^{\dagger}a_{2}^{\dagger}a_{1}a_{2}\rangle- 
\langle a_{1}^{\dagger}a_{1}^{\dagger}a_{2}a_{2}\rangle-
\langle a_{2}^{\dagger}a_{2}^{\dagger}a_{1}a_{1}\rangle)\cos^2\theta \sin^2\theta\nonumber\\
 &+& 2i(\langle a_{1}^{\dagger}a_{1}^{\dagger}a_{1}a_{2}\rangle-
\langle a_{1}^{\dagger} a_{2}^{\dagger} a_{1} a_{1}\rangle)\cos^3\theta \sin\theta\nonumber
\\
&+&2i(\langle a_{1}^{\dagger} a_{2}^{\dagger} a_{2} a_{2} \rangle- 
\langle a_{2}^{\dagger} a_{2}^{\dagger} a_{1} a_{2}
\rangle)\cos\theta\sin^3\theta .
\end{eqnarray}
Here, three suitably chosen values of $\theta$ are sufficient to determine 
the terms with different dependence on $\theta$ 
since the first two terms were determined previously. Finally, choosing 
\begin{equation}
u= \frac{1}{\sqrt{2}}\left(\,
\matrix{ 1 & e^{i\pi/4} \cr
-e^{-i\pi/4}&1}\,\right),
\end{equation}
we obtain
\begin{eqnarray}
\langle b_{1}^{\dagger}(u)b_{2}^\dagger(u)b_{1}(u)b_2(u)\rangle&=&
\frac{1}{4}\langle a_{1}^{\dagger}a_{1}^{\dagger}a_{1}a_{1}\rangle + 
\frac{1}{4}\langle a_{2}^{\dagger}a_{2}^{\dagger}a_{2}a_{2}\rangle\nonumber\\
&-&\frac{i}{4}(\langle a_{1}^{\dagger}a_{1}^{\dagger}a_{2}a_{2}\rangle- 
\langle a_{2}^{\dagger}a_{2}^{\dagger}a_{1}a_{1}\rangle).
\end{eqnarray}
Thus the nine choices of $u$ above permit us to determine the nine field 
correlations listed after Eq.(5) and hence all the variances in the Stokes 
operators. Besides the variances,  the same nine correlations determine the
correlations among Stokes operators. For example the last term in Eq.(15) 
represents the normally ordered correlation $
\left\langle
{\tiny
\matrix{
\circ \cr \circ  
}
}
S_2 S_3
{\tiny 
\matrix{
\circ \cr \circ
}
} \right\rangle.$
Further note that (11) [(13)] for
$\theta=\pi/4$ and $\theta=3\pi/4$ when added and subtracted would yield 
$
\left\langle
{\tiny
\matrix{
\circ \cr \circ  
}
}
S_0 S_0
{\tiny 
\matrix{
\circ \cr \circ
}
} \right\rangle+
\left\langle
{\tiny
\matrix{
\circ \cr \circ  
}
}
S_2 S_2
{\tiny 
\matrix{
\circ \cr \circ
}
} \right\rangle$
 $\bigg[
\left\langle
{\tiny
\matrix{
\circ \cr \circ  
}
}
S_0 S_0
{\tiny 
\matrix{
\circ \cr \circ
}
} \right\rangle$-$
\left\langle
{\tiny
\matrix{
\circ \cr \circ  
}
}
S_2 S_2
{\tiny 
\matrix{
\circ \cr \circ
}
} \right\rangle\bigg]$ and  
$
\left\langle
{\tiny
\matrix{
\circ \cr \circ  
}
}
S_0 S_2
{\tiny 
\matrix{
\circ \cr \circ
}
} \right\rangle$
$
\left[\left\langle
{\tiny
\matrix{
\circ \cr \circ  
}
}
S_0 S_3
{\tiny 
\matrix{
\circ \cr \circ
}
} \right\rangle\right].$

The unitary transformations $u$ involved in the measurements above are all 
of the form 
\begin{equation}
u(\theta,\phi)= \left(\,
\matrix{ \cos\theta & e^{i\phi}\sin\theta \cr
-e^{-i\phi}\sin\theta &\cos\theta  \cr}\,\right),
\end{equation}
with $\phi$ taking values $0, \pi/4, \pi/2$.
These transformations can most conveniently be realized using the universal 
$SU(2)$ gadget\cite{simon1,simon2} which consists of two quarter wave plates 
and one half wave plate mounted coaxially. Taking the axis of the gadget to be 
along the $z$ axis and the slow axes of the quarter and half wave plates in 
the $x$ direction,    
the settings for the $SU(2)$ gadget in the Q-Q-H 
configuration which realize the required transformations are 
\begin{mathletters}
\begin{eqnarray}
u(\theta,0)&=& Q_{\pi/4}~Q_{\pi/4}~H_{-\pi/4+\theta/2},\\
u(\theta,\pi/2)&=& Q_{\pi/2}~Q_{(\theta+\pi/2)}~H_{\theta/2},\\
u(\pi/4, \pi/4)&=& Q_{\pi/4+(\tan^{-1}{\sqrt{2}})/2}
~Q_{5\pi/12+(\tan^{-1}{\sqrt{2}})/2}~H_{\pi/12},
\end{eqnarray}
\end{mathletters}
where the the subscripts to $Q$ and $H$ give the angles by which the quarter 
and the half wave plates have to be rotated around the $z$ axis. 
The matrices for the quarter wave and half wave plates are given by
\begin{mathletters}
\begin{eqnarray}
H_{\phi} &=& i \left(\matrix{ \cos 2\phi & \sin 2\phi \cr
\sin 2\phi & -\cos 2\phi} \right), \\
Q_{\phi} &=&
\frac{i}{\sqrt{2}}\left(\matrix{ \cos 2\phi -i & \sin 2\phi \cr
\sin 2\phi & -\cos 2\phi - i} \right).
\end{eqnarray}
\end{mathletters}
After `rotating' the $a_1$ and $a_2$ into $b_1(u)$ and $b_2(u)$ modes with 
the help of the $SU(2)$ gadget one can separate  the $b_1(u)$, $b_2(u)$ modes 
with the help of a polarising beam splitter and carry out the required 
intensity and intensity-intensity correlation measurements to obtain complete information 
about the Stokes parameters and fluctuations thereof.\\

To conclude,  we have shown how a combination of quarter wave and half wave
plates can be used to measure not only the Stokes parameters but also all the
quantum correlations and fluctuations in the Stokes parameters.


\begin{references}
\bibitem{born} M. Born and E. Wolf,~ {\it Principles of Optics},~ Seventh
Edition, (Cambridge University Press 1999), Sec.1.4.
\bibitem{mandel} L. Mandel and E. Wolf, ``Optical Coherence and Quantum Optics"
(Cambridge University Press, 1995), Chap.6.
\bibitem{wolf} E. Wolf, Nuovo Cimento {\bf 13}, 1165 (1959).
\bibitem{korol} N. V. Korolkova, G. Leuchs, R. Loudon, T. C. Ralph and C. 
Silberhorn, quant-ph/0108098.
\bibitem{chirkin} N. V. Korolkova and A. S. Chirkin, J. Mod. Opt. {\bf 43}, 869
(1996).
\bibitem{abouraddy}.  A. F. Abouraddy,  A. V. Sergienko,  B. E. A. Saleh  and  
M. C. Teich,  Opt. Commun. {\bf 201},  93  (2002);  D. F. V. James,  P. G. 
Kwiat, W. J. Munro and A. G. White,  Phys.  Rev. A {\bf 64},  052312  (2001).
\bibitem{Jauch} J. M. Jauch and F. Rohrlich,  {\it The Theory of Photons and 
Electrons} (Springer-Verlag, Berlin 1980) Sec.2.8.   B.A. Robson,  
{\it The Theory of Polarization Phenomena} (Clarendon Press, Oxford 1974).
\bibitem{schwinger} J. Schwinger, Proc.  Nat.  Acad.  Sci.  U.S.A {\bf 46}, 
570 (1960). 
\bibitem{agarwal} G. S.Agarwal and R. R. Puri, Phys. Rev. A  {\bf 41},  3782  
(1990);
{\it ibid} {\bf 49}, 4968 (1994); A.  Kuzmich,  L. Mandel  and  N. P. Bigelow,  Phys.  Rev.
Lett. {\bf 85}, 1594 (2000), A.  Sorensen  and  K. Molmer  quant-ph/0011035.
\bibitem{grangier}  P.  Grangier,  R. E.  Slusher,  B. Yurke  and  A. LaPorta,
Phys. Rev.  Lett. {\bf 59},  2153  (1987).
\bibitem {puri}  A .P. Alodjants, A. M. Arakelian and A. S. Chirkin,  
Appl. Phys. B
{\bf 66},  53  (1998);  G. S. Agarwal  and  R. R. Puri, Phys.  Rev.  A  
{\bf 40}, 5179  (1989).
\bibitem{mukunda}  N. Mukunda  and  T. F. Jordan,  J. Math.  Phys. {\bf 7}, 
849  (1966).
\bibitem{bowen} W. P. Bowen, R. Schnabel,  Hans A. Bachor  and  P. K. Lam, 
Phys. Rev. Lett.  {\bf 88},  093601  (2002).
\bibitem{walls} D. F. Walls  and  G. J. Milburn,  {\it Quantum  Optics} (Springer-Verlag,
Berlin, 1994); Chap.V.
\bibitem{simon1} R. Simon  and  N. Mukunda,  Phys.  Lett.  A {\bf 138},  474  (1990).
\bibitem{simon2} R. Simon  and  N. Mukunda,  Phys.  Lett.  A {\bf 143},  165  (1990).
\end{references}
\end{document}